# Transient and Steady-State Temperature Rise in Three-Dimensional Anisotropic Layered Structures in Pump-Probe Thermoreflectance Experiments

Puqing Jiang,[1,2,a)] and Heng Ban[1,2,b)]

[1]*Department of Mechanical Engineering and Materials Science, University of Pittsburgh, Pittsburgh, Pennsylvania 15261, USA*

[2]*Pittsburgh Quantum Institute, Pittsburgh, Pennsylvania 15260, USA*

**ABSTRACT**

Recent developments of the pump-probe thermoreflectance methods (such as the beam-offset and elliptical-beam approaches of the time-domain and frequency-domain thermoreflectance techniques) enabled measurements of the thermal conductivities of in-plane anisotropic materials. Estimating the temperature rise of anisotropic layered structures under surface heating is critically important to make sure that the temperature rise is not too high to alias the signals in these experiments. However, a simple formula to estimate the temperature rise in three-dimensional (3D) anisotropic layered systems heated by a non-circular laser beam is not available yet, which is the main problem we aim to solve in this work. We first re-derived general formalisms of the temperature rise of a multilayered structure based on the previous literature work by solving the 3D anisotropic heat diffusion equation in the frequency domain. These general formalisms normally require laborious numerical evaluation; however, they could be reduced to explicit analytical expressions for the case of semi-infinite solids. We then extend the analytical expressions to multilayered systems, taking into account the effect of the top layers. This work not only enhances our understanding of the physics of temperature rise due to surface laser heating but also enables quick estimation of the peak temperature rise of 3D anisotropic layered systems in pump-probe thermoreflectance experiments and thus greatly benefits the thermoreflectance experiments in choosing the appropriate heating power intensity for the experiments.

---

[a] puqing.jiang@pitt.edu
[b] heng.ban@pitt.edu

## I. INTRODUCTION

Thermal conductivity of anisotropic materials is a critical property that impacts their wide applications in modern devices such as microelectronics,[1] photonics,[2] solar cells,[3] thermal barrier coatings[4], and thermoelectric modules.[5] Accurate measurements of the three-dimensional (3D) thermal conductivity of thin films and small-scale samples were made possible with the recent development of transient techniques including the thermoreflectance methods (time-domain/frequency-domain thermoreflectance, TDTR/FDTR) and the 3ω method. These techniques have been extensively used to measure the 3D thermal conductivity tensor of both thin films and bulk materials over a wide range of thermal conductivity.[6, 7 8-21]

The thermoreflectance methods, including both TDTR and FDTR, use a modulated pump laser beam to heat the sample and a second probe beam to detect the surface temperature change via the linear change of the surface reflectance $R_s$ with temperature, $\Delta R_s = \frac{dR_s}{dT}\Delta T$. TDTR typically uses ultrafast pulsed lasers (with pulse widths <0.5 ps) and FDTR typically uses continuous-wave (cw) lasers for the measurements. In both cases, the surface temperature rise $\Delta T$ should not be too high to invalidate the assumption of the linear temperature dependence of $R_s$. Besides, since both methods measure the thermal properties of the sample at the base temperature $T_0$, the surface temperature rise $\Delta T$ should not be too high to invalidate the assumption of constant thermal properties in the temperature range of $T_0$ to $T_0 + \Delta T$. This is particularly important at low temperatures where the heat capacities have the $T^3$ temperature dependence. A general guideline is that the steady-state surface temperature rise $\Delta T$ should not exceed 10 K or 10% of the absolute temperature, whichever is smaller.[6, 22] A quick and accurate estimation of the temperature rise in both TDTR and FDTR is thus critically important in choosing an appropriate heating power intensity for the experiments.

For the steady-state temperature rise in thermoreflectance experiments, Cahill[23] was the first to derive an analytical expression for an isotropic semi-infinite solid, which was later extended to a more general form by Braun *et al.*[24] as

$$\Delta T = \frac{P_0}{\sqrt{2\pi k_z k_r (w_0^2 + w_1^2)}}. \tag{1}$$

Here, $P_0$ is the averaged laser power absorbed by the sample, $k_r$ and $k_z$ are the in-plane and through-plane thermal conductivities of the material, $w_0$ and $w_1$ are the $1/e^2$ radii of the pump and probe laser spots, and $\Delta T$ is the temperature rise as averaged by the probe beam. While Eq. (1) has been commonly used in both TDTR and FDTR experiments to estimate the steady-state temperature rise, it is limited to axially symmetric cases. Besides, the case of pulsed laser heating in TDTR should be more complicated than Eq. (1), and the peak transient temperature rise in TDTR due to the pulsed laser heating could be much higher than the steady-state temperature rise estimated by Eq. (1), especially at low temperatures where the sample could have a very high in-plane thermal diffusivity $k_r/C$. Also, as pointed out by Braun *et al.*[24], the layers and interfaces in multilayered systems could redistribute the heat flux and result in a temperature rise that is different from the one predicted using Eq. (1). A more general formula of the temperature rise for both cw and pulsed laser heating that can apply to multilayered systems with 3D anisotropy is urgently needed.

In this work, general formalisms of the peak temperature rise were first re-derived based on the previous literature work for both cw and pulsed laser heating of 3D anisotropic multilayered systems. These general formalisms were then reduced to simple semi-empirical correlations to enable fast and accurate estimation of the peak temperature rise in both FDTR and TDTR experiments. These simple analytical expressions were tested against the general formalisms for several different representative cases and the effect of thin-film layers on the peak temperature rise was discussed.

## II. GENERAL FORMALISM

The heat diffusion process in anisotropic media is governed by the following equation:

$$C\frac{\partial T}{\partial t} = k_x \frac{\partial^2 T}{\partial x^2} + k_y \frac{\partial^2 T}{\partial y^2} + k_z \frac{\partial^2 T}{\partial z^2} + 2k_{xy}\frac{\partial^2 T}{\partial x \partial y} + 2k_{xz}\frac{\partial^2 T}{\partial x \partial z} + 2k_{yz}\frac{\partial^2 T}{\partial y \partial z}. \qquad (2)$$

Here $C$ is the volumetric heat capacity, and $k_x, k_y, k_z, k_{xy}, k_{yz}, k_{xz}$ are the 6 independent components of the 2nd rank thermal conductivity tensor, which must be symmetric by the reciprocity relation.[25, 26] The temperature dependence of the thermophysical properties is ignored for the sake of simplicity. This parabolic partial differential equation could be simplified to a one-dimensional (1D) heat conduction problem in the frequency domain by doing Fourier transforms to the in-plane coordinates and time, $T(x, y, z, t) \leftrightarrow \Theta(u, v, z, \omega)$. Assuming a surface heat source boundary condition, the surface temperature rise $\Theta_s$ of a multilayered system could be solved in the frequency domain as

$$\Theta_s(u, v, \omega) = \hat{G}(u, v, \omega) Q_s(u, v, \omega), \qquad (3)$$

where $\hat{G}(u, v, \omega)$ is the Green's function and $Q_s(u, v, \omega)$ is the surface heat flux in the frequency domain. Details on the derivation of the Green's function can be found in Refs. [10, 14] and also Supplementary Information Section S1 for the sake of completeness.

We note that the assumption of a surface heat source boundary condition is not always valid due to the optical penetration of the laser beam into the sample. However, for most standard TDTR and FDTR experiments where the samples are coated with a metal transducer layer, most of the heat is absorbed by the thin transducer layer. Assuming the surface heat source boundary condition would only result in slightly higher temperature rise at short delay times (<10 ps) in TDTR but have no effect for the longer delay time range (>10 ps) in TDTR or FDTR. More detailed discussions can be found in Supplementary Information S2. In what follows, we continue to derive the temperature rise expressions under the assumption of a surface heat source boundary condition.

For the case of cw laser heating in FDTR, the surface heat flux is

$$q^{cw}(x,y,t) = \frac{2}{\pi w_x w_y} \exp\left[-2\left(\frac{x^2}{w_x^2} + \frac{y^2}{w_y^2}\right)\right]\left[P_0 + P_1 \cos(\omega_0 t + \varphi_0)\right]. \tag{4}$$

Here $w_x$ and $w_y$ are the $1/e^2$ radii of the pump laser spot in the $x$ and $y$ directions, respectively; $P_0$ and $P_1$ are the amplitudes of the offset component and the periodic heating power [in Watts] absorbed by the sample, with $P_0 \geq P_1$; $\omega_0$ is the modulation frequency, and $\varphi_0$ is an arbitrary phase shift. Fourier transform of Eq. (4) gives its frequency-domain expression as

$$Q_s^{cw}(u,v,\omega) = 2\pi \exp\left(-\frac{\pi^2 u^2 w_x^2}{2}\right)\exp\left(-\frac{\pi^2 v^2 w_y^2}{2}\right) \times \left[P_0\delta(\omega) + P_1 \frac{\delta(\omega-\omega_0)e^{i\varphi_0} + \delta(\omega+\omega_0)e^{-i\varphi_0}}{2}\right]. \tag{5}$$

The case of pulsed laser heating in TDTR would be a little more complicated, with the surface heat flux expressed as

$$q^{pulsed}(x,y,t) = \frac{2}{\pi w_x w_y} \exp\left[-2\left(\frac{x^2}{w_x^2} + \frac{y^2}{w_y^2}\right)\right]\left[P_0 + P_1 \cos(\omega_0 t + \varphi_0)\right] \times \left[\frac{0.94}{f_{rep}\tau_p} \sum_{n=-\infty}^{\infty} \exp\left(-2.77\frac{(t-nT_s-t_0)^2}{\tau_p^2}\right)\right]. \tag{6}$$

This is a train of pulses with Gaussian distribution in time and space, modulated by a sinusoidal function at frequency $\omega_0$. The laser pulses have a full duration at half maximum (FDHM) of $\tau_p$. The laser repetition rate is $f_{rep}$ with a period of $T_s=1/f_{rep}$, and $t_0$ is an arbitrary time shift of laser pulses. Similarly, the frequency-domain surface heat flux for the case of pulse laser heating could be obtained from the Fourier transform of Eq. (6) as:

$$Q_s^{pulsed}(u,v,\omega) = 2\pi \exp\left(-\frac{\pi^2 u^2 w_x^2}{2}\right)\exp\left(-\frac{\pi^2 v^2 w_y^2}{2}\right)\times$$

$$\sum_{n=-\infty}^{\infty}\left\{\left[P_0\delta(\omega-n\omega_s)+P_1\frac{\delta(\omega-n\omega_s-\omega_0)e^{i\varphi_0}+\delta(\omega-n\omega_s+\omega_0)e^{-i\varphi_0}}{2}\right]e^{-\frac{n^2\omega_s^2\tau_p^2}{11.08}}e^{-in\omega_s t_0}\right\}.$$

(7)

The surface temperature rise of the layered structure can thus be obtained by doing an inverse 2D Fourier transform of Eq. (3) as

$$\theta(x,y,t) = \frac{1}{2\pi}\int_{-\infty}^{\infty}\int_{-\infty}^{\infty}\int_{-\infty}^{\infty} Q_s(u,v,\omega)\hat{G}(u,v,\omega)\exp(i2\pi(ux+vy)+i\omega t)\,du\,dv\,d\omega. \tag{8}$$

For the case of cw laser heating, the time-varying surface temperature rise is

$$\theta(x,y,t) = P_0\int_{-\infty}^{\infty}\int_{-\infty}^{\infty}\hat{G}(u,v,0)\exp\left(-\frac{\pi^2 u^2 w_x^2}{2}\right)\exp\left(-\frac{\pi^2 v^2 w_y^2}{2}\right)\exp(i2\pi(ux+vy))\,du\,dv$$

$$+P_1\int_{-\infty}^{\infty}\int_{-\infty}^{\infty}\mathrm{Re}\left\{\hat{G}(u,v,\omega_0)e^{i(\omega_0 t+\varphi_0)}\right\}\exp\left(-\frac{\pi^2 u^2 w_x^2}{2}\right)\exp\left(-\frac{\pi^2 v^2 w_y^2}{2}\right)\exp(i2\pi(ux+vy))\,du\,dv$$

(9)

Here the symbol Re{ } represents the real part of a complex number.

For the case of pulsed laser heating, it would be

$$\theta(x,y,t) = P_0\sum_{n=-\infty}^{\infty}\left\{e^{-\frac{n^2\omega_s^2\tau_p^2}{11.08}}e^{in\omega_s(t-t_0)}\int_{-\infty}^{\infty}\int_{-\infty}^{\infty}\hat{G}(u,v,n\omega_s)\exp\left(-\frac{\pi^2 u^2 w_x^2}{2}\right)\exp\left(-\frac{\pi^2 v^2 w_y^2}{2}\right)\exp(i2\pi(ux+vy))\,du\,dv\right\}$$

$$+P_1\sum_{n=-\infty}^{\infty}\left\{e^{-\frac{n^2\omega_s^2\tau_p^2}{11.08}}e^{in\omega_s(t-t_0)}\int_{-\infty}^{\infty}\int_{-\infty}^{\infty}\mathrm{Re}\left\{\hat{G}(u,v,n\omega_s+\omega_0)e^{i\omega_0 t}\right\}\exp\left(-\frac{\pi^2 u^2 w_x^2}{2}\right)\exp\left(-\frac{\pi^2 v^2 w_y^2}{2}\right)\exp(i2\pi(ux+vy))\,du\,dv\right\}$$

(10)

The first terms of Eqs. (9) and (10) are the steady-state temperature rise induced by the constant offset component of the heat flux, and the second terms of Eqs. (9) and (10) are the steady periodic temperature oscillation induced by the modulated component of the surface heat flux.

Figure 1 (a, b) shows an example of the time-varying heating power of cw and pulsed laser heating with the same averaged heating power of $P_0 = 10$ mW and the same amplitude of the periodic heating power as $P_1 = 10$ mW. Here the pulsed laser has a repetition rate of $f_{rep} = 80$ MHz and a pulse width of $\tau_p = 0.5$ ps. Figure 1(b) shows that although the averaged power of the pulsed laser is only 10 mW, the instant laser power could be as high as 300 W due to the ultrashort widths of the laser pulses.

Figure 1 (c-e) shows the corresponding peak temperature rises (located at $x = y = z = 0$) of a 100 nm Al/Si two-layered system as a function of time under both cw and pulsed laser heating, whose power profiles are shown in Fig. 1 (a, b). The subplots (1c-1e) are for the system at 300 K (with $k_{Al} = 200$ W/mK, $C_{Al} = 2.34$ MJ/m³K, $k_{Si} = 140$ W/mK, $C_{Si} = 1.6$ MJ/m³K) and the subplots (2c-2e) are for the system at 70 K (with $k_{Al} = 80$ W/mK, $C_{Al} = 0.77$ MJ/m³K, $k_{Si} = 1680$ W/mK, $C_{Si} = 0.35$ MJ/m³K). The laser spot size was fixed as $w_x = w_y = 5$ μm and the modulation frequency was chosen as 5 MHz for both cases. The blue curves in Fig. 1 are for the pulsed laser heating and the red shadowed regions represent the cw laser heating. We note that in TDTR experiments the signals in the short delay time range 0-80 ps are usually excluded from being fitted by the thermal model prediction because data in this delay time range are complicated by heat transport processes in the transducer layer and are not important for extracting thermal properties of the substrate. Therefore, for the time-varying temperature rise of pulsed laser heating, the data in the delay time range 0-80 ps after each pulse were omitted for the sake of clarity.

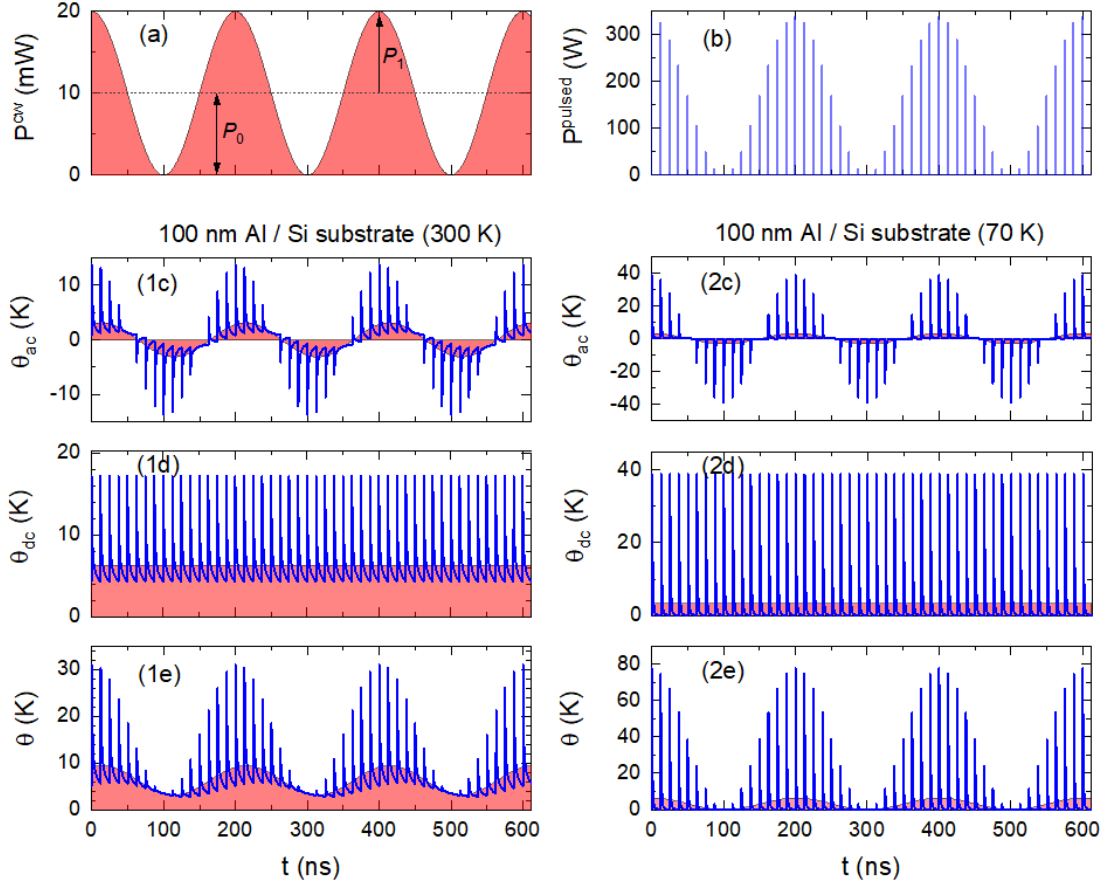

FIG. 1. (a, b) The time-varying heating power profile of modulated cw and pulsed laser with the same averaged heating power of $P_0 = 10$ mW and the same amplitude of oscillating heating power of $P_1 = 10$ mW. (c-e) The peak temperature rise (located at $x = y = z = 0$) $\theta_{ac}$, $\theta_{dc}$, and $\theta = \theta_{ac} + \theta_{dc}$ of a 100 nm Al/Si system as a function of time at different temperatures of 300 K (subplots 1c – 1e) and 70 K (subplots 2c – 2e) heated by both a pulsed laser (the blue curves) and a cw laser (the red shadowed region). The heating power, laser spot size, and modulation frequency are all kept the same for both cases. For the case of pulsed laser heating, the data in the delay time range 0-80 ps after each pulse were omitted for the sake of clarity.

As shown in Fig. 1 (c-e), the temperature rise of the sample consists of a steady-state component $\theta_{dc}$ induced by the constant offset component of the surface heat flux and a periodic oscillating component $\theta_{ac}$ induced by the modulated component of the surface heat flux. The amplitude of $\theta_{ac}$ is always smaller than or equal to that of $\theta_{dc}$. For the pulsed laser heating, the temperature rise consists of a pulse accumulation component, which has an amplitude similar to that of the cw laser heating, and a transient component induced by each pulse heating event, whose amplitude decreases approximately exponentially with the delay time after the pulse heating event. The transient component of the temperature rise at a short delay time could be much higher than that of the cw laser heating, especially at low temperatures where the sample has a high in-plane thermal diffusivity $k_r/C$. For example, for the case shown in Fig. 1 (2e), Eq. (1) estimates a steady-state temperature rise of ~0.47 K for the Si substrate. However, the transient temperature rise in TDTR at a delay time of 80 ps would vary in the range 0-80 K, as shown in Fig. 1 (2e). The commonly used rule of thumb that "the steady-state temperature rise should not exceed 10 K or 10% of the absolute temperature" [6, 22] is thus not valid here. Therefore, it would be insufficient to only estimate the steady-state temperature rise based on Eq. (1) in TDTR experiments. In what follows, we will develop simple analytical expressions for quick estimation of the full temperature rise for both cw and pulsed laser heating. These analytical expressions also help us identify the key parameters affecting the temperature rise and better understand the correlations among them.

**III. ANALYTICAL EXPRESSIONS FOR THE PEAK TEMPERATURE RISE OF SEMI-INFINITE SOLIDS**

The general formalisms derived above (Eqs. (9) and (10)) could be reduced to explicit analytical expressions for the simple case of a semi-infinite solid with an orthogonal thermal conductivity tensor ($k_{xy} = k_{xz} = k_{yz} = 0$), for which the Green's function is

$$\hat{G}(u,v,\omega) = \frac{1}{\sqrt{iC\omega k_z + 4\pi^2 k_z \left(k_x u^2 + k_y v^2\right)}} . \tag{11}$$

*3.1 Peak temperature rise of a semi-infinite solid induced by cw laser heating*

Plugging Eq. (11) into Eq. (9) and setting $x = y = 0$, the spatially peak temperature rise of a semi-infinite solid under cw laser heating is expressed as:

$$\theta^{cw} = P_0 \int_{-\infty}^{\infty}\int_{-\infty}^{\infty} \frac{\exp(-\pi^2 u^2 w_x^2/2)\exp(-\pi^2 v^2 w_y^2/2)}{2\pi\sqrt{k_z(k_x u^2 + k_y v^2)}} dudv$$

$$+ P_1 \operatorname{Re}\left\{ e^{i\omega_0 t} \int_{-\infty}^{\infty}\int_{-\infty}^{\infty} \frac{\exp(-\pi^2 u^2 w_x^2/2)\exp(-\pi^2 v^2 w_y^2/2)}{\sqrt{iC\omega_0 k_z + 4\pi^2 k_z(k_x u^2 + k_y v^2)}} dudv \right\}. \tag{12}$$

The first term of Eq. (12), which is due to the constant offset component of the cw laser heating, can be expressed as

$$\theta^{cw}_{dc} = \frac{P_0}{\sqrt{2\pi w_0^2 k_z k_r}} \xi, \tag{13}$$

with the anisotropy correction factor $\xi(\frac{\alpha}{\beta})$ expressed as

$$\xi\left(\frac{\alpha}{\beta}\right) = \frac{w_0}{\sqrt{2\pi}} \int_{-\infty}^{\infty}\int_{-\infty}^{\infty} \frac{\exp[-\pi^2 w_0^2(u^2/\beta + \beta v^2)/2]}{\sqrt{(u^2/\alpha + \alpha v^2)}} dudv. \tag{14}$$

Here the anisotropy parameters are defined as $\alpha = \sqrt{k_y/k_x}$, $\beta = w_y/w_x$, and the averaged in-plane thermal conductivity and laser spot size are defined as $k_r = \sqrt{k_x k_y}$ and $w_0 = \sqrt{w_x w_y}$, respectively.

The anisotropy correction factor $\xi(\alpha/\beta)$ is a function of solely $\alpha/\beta$ and is plotted out as the solid curve in Fig. 2. It could also be approximated as $\xi(\alpha/\beta) \approx \frac{4\sqrt{\alpha\beta}}{(\sqrt{\alpha}+\sqrt{\beta})^2}$, which has an error of <2% if $\alpha/\beta$ is in the range $0.1 < \alpha/\beta < 10$ and an error of ~10% if $0.01 < \alpha/\beta < 100$, as shown by the dashed curve in Fig. 2. If the anisotropy is too large with $\alpha/\beta > 100$ or $\alpha/\beta < 0.01$, a slightly different correlation $\xi(\alpha/\beta) \approx \frac{4(\frac{\alpha}{\beta})^{0.45}}{\left(1+(\frac{\alpha}{\beta})^{0.45}\right)^2}$ could be used instead to approximate $\xi$ with an error <5%, as shown by the dash-dotted curve in Fig. 2.

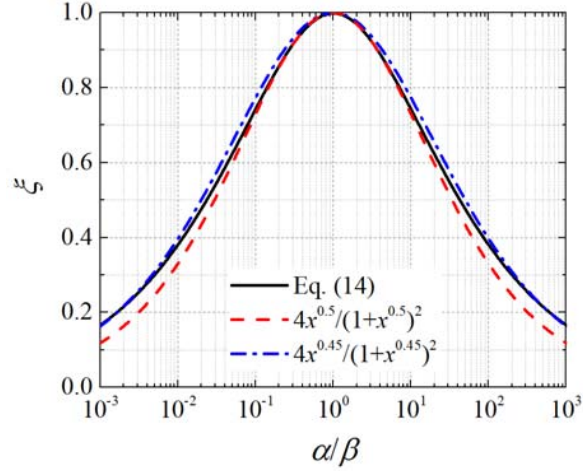

FIG. 2. The anisotropy correction factor $\xi$ (Eq. (14)) as a function of $\alpha/\beta$, with $\alpha = \sqrt{k_y/k_x}$ and $\beta = w_y/w_x$. The solid curve is the accurate numerical result of Eq. (14). The dashed and dash-dotted curves are the approximations.

When $\alpha = \beta$ which gives $\xi = 1$, Eq. (13) becomes identical to Eq. (1) if the probe laser spot size $w_1$ in Eq. (1) is set to be 0. Note that the temperature rise derived here is the peak value induced by the pump heating and is independent of the probe spot size, whereas the temperature rise derived by Cahill[23] and Braun *et al.*[24] was the value averaged by the Gaussian profile of the probe beam. In the limit $w_1 \rightarrow 0$, the probe beam detects only the peak temperature induced by the pump heating. In what follows, we focus only on the peak temperature rise induced by the pump beam; therefore, the laser spot sizes $w_x$ and $w_y$ in the correlations represent the $1/e^2$ radii of only the pump laser spot in the $x$ and $y$ directions, respectively. If the probe-averaged temperature rise is desired instead, simply set $w_x = \sqrt{w_{x_0}^2 + w_{x_1}^2}$ and $w_y = \sqrt{w_{y_0}^2 + w_{y_1}^2}$ in the correlations, where $w_{x_0}$ and $w_{y_0}$ are for the pump laser spot and $w_{x_1}$ and $w_{y_1}$ are for the probe laser spot.

Equation (13) suggests that the steady-state temperature rise of the semi-infinite solid is proportional to both the heat flux and the square-root of the heated area, $\theta \propto \frac{q''}{k}\sqrt{A}$, where $A$ is the heated area and $k = \sqrt{k_z k_r}$ is the averaged thermal conductivity. Therefore, even under the continuous heating of a constant heat flux without any heat loss, a steady-state temperature rise could still be established in the semi-infinite solid if the heated area is finite. On the other hand, for the case of 1D uniform heating with an infinitely large heated area, the surface temperature rise would be $\theta = \frac{2q''}{k}\sqrt{\frac{kt}{\pi C}} \propto \sqrt{t}$, where $t$ is the heating time,[27] in which case a steady state could never be established.

The second term of Eq. (12), which is due to the heating by the modulated component of the cw laser heat flux, is oscillating with time with a constant amplitude in the fully established state. To understand how the heating frequency affects the amplitude of the steady periodic temperature oscillation, we plot in Fig. 3 the ratios between the amplitudes of the periodic temperature oscillation at different heating frequencies and the steady-state temperature rise as solid curves, assuming $P_0 = P_1$. We find that the temperature rise ratios mainly depend on the normalized laser spot size $\hat{w} = w_0/\sqrt{\frac{2\pi k_r}{\omega_0 C}}$, where $w_0$ and $k_r$ are the averaged laser spot size and in-plane thermal conductivity defined as $w_0 = \sqrt{w_x w_y}$ and $k_r = \sqrt{k_x k_y}$, respectively. The amplitude of the steady periodic temperature oscillation of the semi-infinite solid is always less than or equal to its steady-state temperature rise under the same amplitude of heating power. When $w_0 \ll \sqrt{\frac{2\pi k_r}{\omega_0 C}}$ where the heat flow is highly three-dimensional, the amplitude of the periodic temperature oscillation approaches the steady-state temperature rise. On the other hand, with $w_0 \gg \sqrt{\frac{2\pi k_r}{\omega_0 C}}$ the heat flow would be one-dimensional in the $z$-direction; in this case, the amplitude of the periodic temperature oscillation approaches zero.

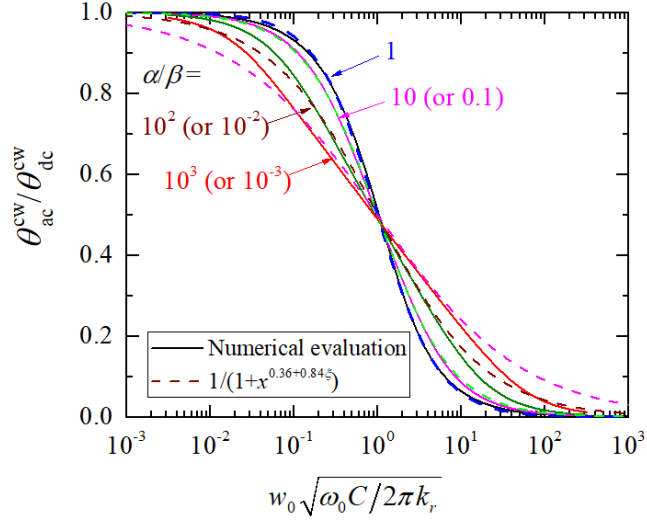

FIG. 3. The ratio of the amplitude of the periodic temperature oscillation to the steady-state temperature rise of a semi-infinite solid plotted as a function of normalized laser spot size for different anisotropy ratios $\alpha/\beta$. The solid curves are the accurate numerical results and the dashed curves are the empirical correlation predictions (Eq. (15)).

The in-plane anisotropy in both the thermal conductivity $\alpha = \sqrt{k_y/k_x}$ and laser spot size $\beta = w_y/w_x$ also affect the amplitude of periodic temperature oscillation, as shown in Fig. 3. We come up with an empirical correlation for the ratio between the amplitude of the periodic temperature oscillation at the frequency $\omega_0$ and the steady-state temperature rise of a semi-infinite solid under cw laser heating as

$$\theta_{ac}^{cw}/\theta_{dc}^{cw} = \frac{1}{1+\left(w_0\sqrt{\dfrac{\omega_0 C}{2\pi k_r}}\right)^{0.36+0.84\xi}}. \qquad (15)$$

Predictions by the correlation Eq. (15) are plotted as the dashed curves in Fig. 3 and they compare quite well with the accurate numerical results if $0.01 < \alpha/\beta < 100$.

In sum, the peak temperature rise of a semi-infinite solid under modulated cw laser heating could be estimated as

$$\theta^{\text{cw}} = \xi \frac{1}{\sqrt{2\pi w_0^2 k_z k_r}} \left\{ P_0 \pm P_1 \left[ 1 + \left( w_0 \sqrt{\frac{\omega_0 C}{2\pi k_r}} \right)^{0.36+0.84\xi} \right]^{-1} \right\}. \qquad (16)$$

The $\pm$ sign represents the range of the temperature oscillation due to the modulated heating.

Note that the steady-state temperature rise due to cw laser heating depends only on the averaged thermal conductivity $k = \sqrt{k_z k_r}$ but not the heat capacity of the material. Although the oscillating part of the temperature rise depends on the heat capacity, the amplitude of the oscillating part of the temperature rise is always less than or equal to the steady-state temperature rise. Therefore, the heat capacity of the material does not place a constraint on the temperature rise in FDTR experiments. The situation, however, is different in TDTR experiments, where a pulsed laser is used for the heating and the heat capacity could play a significant role in the transient temperature rise, see details below.

*3.2 Peak temperature rise of a semi-infinite solid induced by pulsed laser heating*

Plugging Eq. (11) into Eq. (10) and setting $x = y = 0$, the spatially peak temperature rise of a semi-infinite solid under the heating of a train of modulated laser pulses is expressed as:

$$\begin{aligned}\theta^{\text{pulsed}} = &P_0 \sum_{n=-\infty}^{\infty} \left[ e^{-\frac{n^2 \omega_s^2 \tau_p^2}{11.08}} e^{in\omega_s(t-t_0)} \int_{-\infty}^{\infty}\int_{-\infty}^{\infty} \frac{\exp\left(-\pi^2 u^2 w_x^2/2\right)\exp\left(-\pi^2 v^2 w_y^2/2\right)}{\sqrt{iC(n\omega_s)k_z + 4\pi^2 k_z \left(k_x u^2 + k_y v^2\right)}} dudv \right] \\ &+P_1 \text{Re}\left\{ \sum_{n=-\infty}^{\infty} \left[ e^{-\frac{n^2 \omega_s^2 \tau_p^2}{11.08}} e^{in\omega_s(t-t_0)} e^{i\omega_0 t} \int_{-\infty}^{\infty}\int_{-\infty}^{\infty} \frac{\exp\left(-\pi^2 u^2 w_x^2/2\right)\exp\left(-\pi^2 v^2 w_y^2/2\right)}{\sqrt{iC(\omega_0 + n\omega_s)k_z + 4\pi^2 k_z \left(k_x u^2 + k_y v^2\right)}} dudv \right] \right\}\end{aligned} \quad (17)$$

The first term of Eq. (17) is due to the offset component of the pulsed laser heating, an example of which is shown as the blue curves in Fig. 1 (1d, 2d). This temperature rise comprises a constant component due to the pulse accumulation and a transient component due to the heating by each pulse. If the pulse repetition rate $f_{\text{rep}} \to \infty$, the pulse accumulation would approach the cw laser heating and the transient temperature rise due to each pulse would approach zero. Although Eq. (17) is hard to be reduced further but requires numerical evaluation, after analyzing the numerical results of some representative cases, we come up with the following semi-empirical correlation for the first term of Eq. (17) as

$$\theta_{dc}^{pulsed} = \xi \frac{P_0}{\sqrt{2\pi w_0^2 k_z k_r}} \left[1 + \left(16 \frac{k_r}{\pi w_0^2 C} \frac{1}{f_{rep}}\right)^{0.18+0.42\xi}\right]^{-1} + \frac{1.1 P_0}{f_{rep} \pi w_0^2 \sqrt{k_z C t_d}} \left(1 + \frac{25}{\xi^2} \frac{k_r}{\pi w_0^2 C} t_d\right)^{-1}. \quad (18)$$

Here $t_d$ is the time after each pulse heating event. More details on deriving this semi-empirical correlation can be found in Supplementary Information Section S3.

The first term of Eq. (18) is the pulse accumulation $\theta_{dc,accum}^{pulsed}$ and the second term is the transient component $\theta_{dc,trans}^{pulsed}$ due to the individual pulse heating events. The empirical correlation Eq. (18) has been tested against 1000 randomly chosen cases, with $k_x, k_y, k_z$ independently and randomly varying in the range of 0.5-5000 W/mK, $C$ varying in the range of 0.001-10 MJ/m³K, and $w_x, w_y$ independently and randomly varying in the range of 1-100 µm. The tested anisotropy $k_z/k_r$ and $\alpha/\beta$ have covered a wide range from $10^{-4}$ to $10^4$. This correlation was found to be accurate with an error of <3% for $t_d = 100$ ps if the dimensionless factor $\xi w_0 \sqrt{C f_{rep}/k_r} > 3$, which could be met in most cases in TDTR experiments. If $\xi w_0 \sqrt{C f_{rep}/k_r} < 3$ or if $t_d$ is too large (e.g., $t_d > 1$ ns) so that the three-dimensional heat flow dominates, the simple expression of Eq. (18) would cause a larger error on some special occasions but is still generally within 50% of the target values in estimating the temperature rise due to the pulsed heating, see Supplementary Information Section S4 for more details on the validation of this correlation.

From Eq. (18), we can see that the steady-state temperature rise due to pulse accumulation mainly depends on $\frac{q''}{k}\sqrt{A}$, and the transient temperature rise mainly depends on $\frac{q''}{e_z \sqrt{t_d}}$ in the short $t_d$ range, where $e_z = \sqrt{k_z C}$ is the through-plane thermal effusivity. The ratio $w_0/\sqrt{k_r/f_{rep}^2 t_d C}$ is the key factor that determines the relative magnitudes of the pulse-accumulation temperature rise and the transient pulse temperature rise: with $w_0 \gg \sqrt{k_r/f_{rep}^2 t_d C}$ the pulse-accumulation temperature rise would dominate over the transient pulse temperature rise; on the other hand, if $w_0 \ll \sqrt{k_r/f_{rep}^2 t_d C}$, the transient pulse temperature rise could be much higher than the pulse-accumulation temperature rise. A more detailed proof of this statement can be found in Supplementary Information Section S5. Therefore, only estimating the steady-state temperature rise would be insufficient in TDTR experiments especially when measuring materials with high in-plane thermal diffusivities $k_r/C$, an example of which is shown in Fig. 1 (2d).

The second term of Eq. (17) is more complicated since the modulation frequency $\omega_0$ is non-zero. For the modulated pulsed laser heating, the temperature rise $\theta_{ac}^{pulsed}$ also contains a pulse accumulation component $\theta_{ac,accum}^{pulsed}$ and a transient component due to the pulse heating event $\theta_{ac,trans}^{pulsed}$, both periodically varying with time, an example of which is shown as the blue curves in Fig. 1 (1c, 2c). After analyzing the numerical results of 7000 randomly chosen cases with the modulation frequency varying in the range 10 Hz – 10 MHz, we recommend the following empirical correlations to estimate the amplitudes of $\theta_{ac}^{pulsed}$:

$$\begin{aligned}\theta_{ac}^{pulsed} &= \frac{P_1}{P_0}\theta_{dc,accum}^{pulsed}\left[1+4\left(\frac{w_0}{5}\sqrt{\frac{\omega_0 C}{2\pi k_r}}\right)^{0.36+0.84\xi}\right]^{-1}\left[1+\pi\sqrt{\frac{\omega_0}{2\pi f_{rep}}}\right]^{-1} \\ &+\frac{P_1}{P_0}\theta_{dc,trans}^{pulsed}\left[1-\left(1+\frac{(\omega_0/2\pi)^{-0.1\log_{10}(4t_d f_{rep})}}{-\log_{10}(4t_d f_{rep})}\right)^{-1}\left(1+150\left(10w_0\sqrt{\frac{\omega_0 C}{2\pi k_r}}\right)^{-0.6-1.4\xi}\right)^{-1}\right].\end{aligned} \quad (19)$$

The temperature oscillating range by the pulsed heating could thus be estimated as $\theta^{pulsed} = \theta_{dc}^{pulsed} \pm \theta_{ac}^{pulsed}$. The validity of Eq. (19) has been tested against the 7000 random cases, see more details in Supplementary Information Section S6.

As a demonstration, we used the analytical expressions Eqs. (16, 18, 19) to estimate the peak temperature rise of a Si semi-infinite substrate heated at 300 K and 70 K and compare them in Fig. 4 with the accurate numerical evaluations. The laser power was assumed to be absorbed on the surface of the Si substrate. This assumption is necessary here only to evaluate the analytical expressions but is admittedly not physically sound for Si, considering the very small extinction coefficient of Si at most wavelengths. More related discussions can be found in Supplementary Information S2. For the pulsed heating, the temperature rise was evaluated at a delay time $t_d$ of 80 ps. All the other conditions are the same as those for Fig. 1. The shadowed regions in Fig. 4 represent the estimations using the analytical expressions and the curves are the numerical evaluations. The excellent agreement between the analytical estimations and accurate numerical evaluations proves the validity of these analytical expressions.

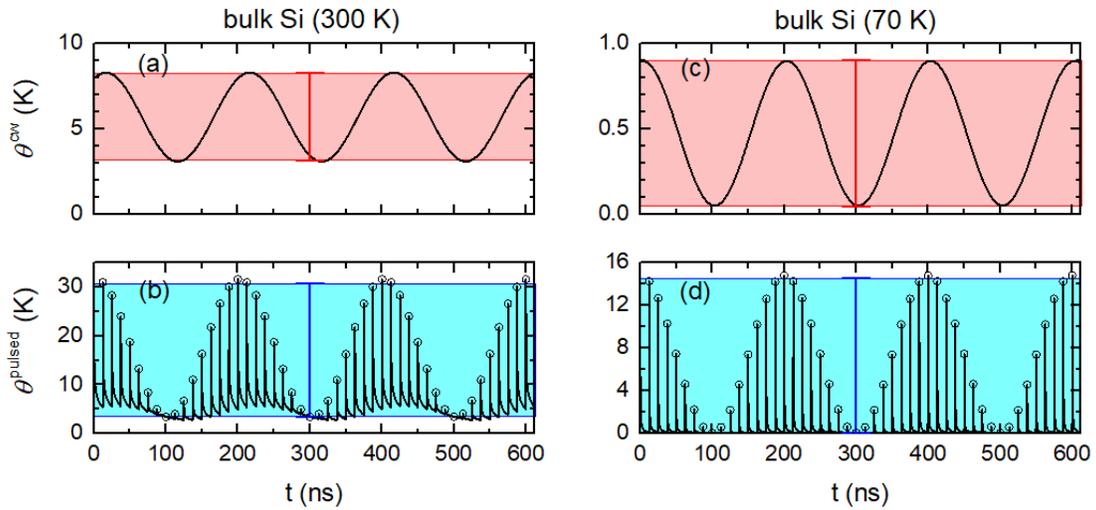

FIG. 4. The time-varying peak temperature rise of a Si substrate heated by a modulated cw laser (a, c) and a modulated pulsed laser (b, d) on the surface at different temperatures of 300 K (a, b) and 70 K (c, d). The averaged heating power is 10 mW and the laser spot size is 5 µm for all the cases. The curves are the numerical results and the shadowed regions represent the estimations using the analytical expressions. The symbols in (b, d) represent the numerical values of the temperature rise evaluated at $t_d = 80$ ps after each pulse.

It would be interesting to compare the temperature rises in Fig. 4 to those in Fig. 1 (1e, 2e). Both temperature rises are evaluated under the same conditions, except that the system for Fig. 1 is 100 nm Al/Si whereas the system for Fig. 4 is a bare Si substrate. Comparison between Fig. 4 and Fig. 1 (1e, 2e) shows that the 100 nm Al layer on top of the Si substrate has little effect on the peak temperature rise of the system at 300 K but would increase the peak temperature rise by ~5 times at 70 K. The top layers could thus have different effects on the peak temperature rise of multilayered systems in different situations and should be carefully evaluated to accurately estimate the peak temperature rise of multilayered systems.

## V. EFFECT OF TOP LAYERS ON THE PEAK TEMPERATURE RISE IN LAYERED STRUCTURES

The case of multilayered systems is more complicated than the semi-infinite solid mainly because the thin films could act as in-plane heat spreaders, making the heat flux intensity on the surface of the underlying layers different from the one originally supplied on the surface of the first layer.

The in-plane heat spreading effect would be larger if the films have a larger $k_r h$, where $h$ is the film thickness. Assuming a linear temperature gradient across the thin films (which could overestimate the temperature rise if the films are too thick to invalidate this assumption), the peak temperature rise of an $n$-layered system could be estimated as:

$$\theta_{n\text{-layered}} = \left[ \frac{2P_0}{\pi w_x w_y} \sum_{i=1}^{n-1}\left(\frac{h_i}{k_{z_i}} + \frac{1}{G_i}\right) + \left(1 + \frac{\sum_{i=1}^{n-1} k_{r_i} h_i}{\sqrt{k_{r_n} k_{z_n}} \sqrt{w_x w_y}}\right)^{-1} \theta_{\text{sub,dc}} \right]$$
$$\pm \left[ \frac{2P_1}{\pi w_x w_y} \sum_{i=1}^{n-1}\left(\frac{h_i}{k_{z_i}} + \frac{1}{G_i}\right) + \left(1 + \frac{\sum_{i=1}^{n-1} k_{r_i} h_i}{\sqrt{k_{r_n} k_{z_n}} \sqrt{w_x w_y}}\right)^{-1} \theta_{\text{sub,ac}} \right], \quad (20)$$

where $h_i$ is the thickness of the $i$-th layer, $G_i$ is the interface thermal conductance between the $i$-th and the next layers, $\theta_{\text{sub,dc}}$ and $\theta_{\text{sub,ac}}$ are the temperature rise of the semi-infinite substrate evaluated using Eqs. (16, 18, 19).

As a demonstration of Eq. (20), we use this equation to calculate the peak temperature rise of two representative three-layered systems: 1) 100 nm Al/50 nm SiO$_2$/Si substrate, and 2) 100 nm Ti/200 nm Si/SiO$_2$ substrate, for both modulated cw and pulsed laser heating. In the first system, the middle layer has a thermal conductivity much lower than the substrate and therefore could incur a significant amount of temperature drop across the film. In the second system, the middle layer has a thermal conductivity much higher than the substrate and therefore should act as an in-plane heat spreader, significantly reducing the heat flux intensity on the surface of the substrate. In the evaluation, the heating power was fixed as $P_0 = P_1 = 10$ mW and an elliptical laser spot with $w_x = 5$ μm and $w_y = 10$ μm was used for both cases. The interface thermal conductance was assumed as $G = 300$ MW/m$^2$K for all the interfaces. The thermal conductivities and heat capacities of all the layers used in the calculation are summarized in Supplementary Information Table S1 in Section S7. The estimated temperature rises are compared with the numerical solutions in Fig. 5, where the shadowed regions represent the estimations using the analytical expressions, and the curves are the accurate numerical evaluation of the time-varying peak temperature rises of the systems. The excellent agreement between our estimations and the accurate numerical results, as shown in Fig. 5, proves the validity of our semi-empirical correlation Eq. (20). This correlation generally works well but could overestimate the temperature rise when the films are too thick to invalidate the assumption of a linear temperature gradient across the films, for example, in the case of Fig. 5 (a, b).

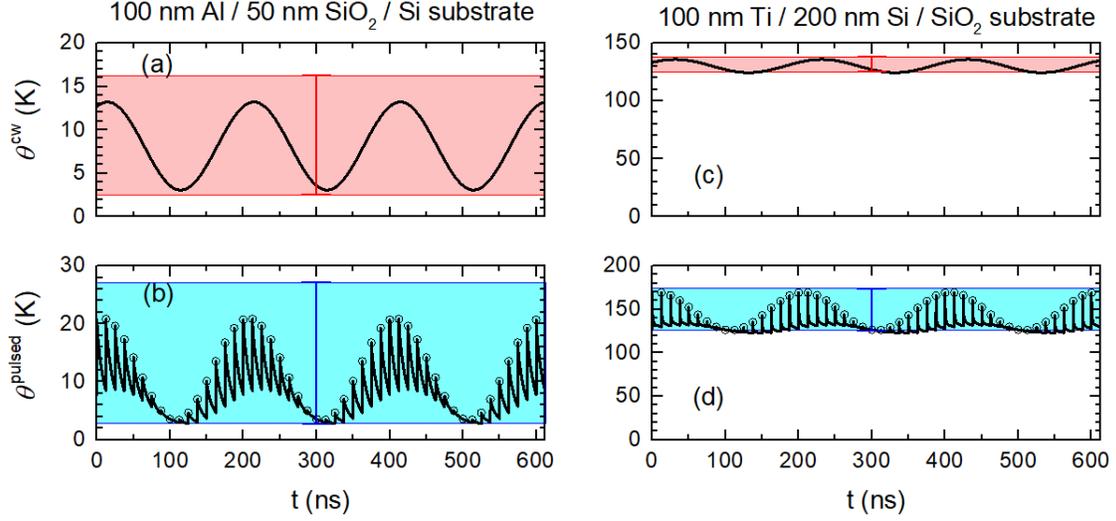

FIG. 5. The time-varying peak temperature rise of a 100 nm Al/50 nm SiO$_2$/Si substrate (a, b) and a 100 nm Ti/200 nm Si/SiO$_2$ substrate (c, d) heated by a modulated cw laser (a, c) and a modulated pulsed laser (b, d). The averaged heating power is 10 mW and the laser spot size is $w_x = 5$ μm and $w_y = 10$ μm for all the cases. The curves are the accurate numerical results and the shadowed regions represent the estimations using the analytical expressions. The symbols in (b, d) represent the numerical values of the temperature rise evaluated at $t_d = 80$ ps after each pulse.

## V. CONCLUSIONS

In summary, we provide simple semi-empirical correlations for quick and accurate estimation of both the steady-state and transient temperature rise of 3D anisotropic multilayered systems heated by either a cw or a pulsed laser beam. General formalisms of the time-varying temperature rise of 3D anisotropic multilayered systems were first re-derived based on the previous literature work for both modulated cw and pulsed laser heating, the evaluation of which, however, requires laborious numerical calculations. We then extract simple semi-empirical correlations of the peak temperature rise of semi-infinite solids from the accurate numerical values of several thousand random cases. These correlations were further extended to multilayered systems, taking into account both the temperature drop across the thin films and their in-plane heat spreading effect. These semi-empirical correlations have been found to work well within the assumption of a linear temperature gradient across the thin films; otherwise, the temperature rise could be overestimated. These simple semi-empirical expressions not only serve the purpose of estimating the maximum temperature rise of 3D anisotropic layered systems in pump-probe thermoreflectance experiments and thus greatly benefits these experiments in choosing the appropriate heating power intensity for the experiments but also enhances our understanding of the physics of temperature rise of multilayered systems under both cw and pulsed laser heating.

## SUPPLEMENTARY MATERIAL

See supplementary material for the derivation of the Green's function, the discussion on the effect of optical penetration, the derivation of Eq. (18), the validation of correlations Eqs. (18, 19), and a table of properties for the cases in Fig. 5.

## ACKNOWLEDGMENTS

P.J. acknowledges the financial support by the R.K. Mellon Postdoctoral Fellowship.

## DATA AVAILABILITY

The data that support the findings of this study are available from the corresponding author upon reasonable request.

Supplementary Information:

Transient and Steady-State Temperature Rise in Three-Dimensional Anisotropic Layered Structures in Pump-Probe Thermoreflectance Experiments

Puqing Jiang,[1,2,a)] and Heng Ban[1,2,b)]

[1]*Department of Mechanical Engineering and Materials Science, University of Pittsburgh, Pittsburgh, Pennsylvania 15261, USA*

[2]*Pittsburgh Quantum Institute, Pittsburgh, Pennsylvania 15260, USA*

**Contents:**

S1: Deriving the Green's function of 3D anisotropic multilayered systems

S2: Effect of optical penetration on the surface temperature rise

S3: Deriving the correlation (Eq. (18)) for $\theta_{\text{dc}}^{\text{pulsed}}$

S4: Validation of the correlation (Eq. (18)) for $\theta_{\text{dc}}^{\text{pulsed}}$

S5: The key factor determining the relative amplitudes of $\theta_{\text{dc,accum}}^{\text{pulsed}}$ and $\theta_{\text{dc,trans}}^{\text{pulsed}}$

S6: Validation of the correlation (Eq. (19)) for $\theta_{\text{ac}}^{\text{pulsed}}$

S7: A table of parameters for the cases in Fig. 5 in the main text

[a] puqing.jiang@pitt.edu
[b] heng.ban@pitt.edu

# S1: Deriving the Green's function of 3D anisotropic multilayered systems

We start from the governing equation of heat diffusion in a 3D anisotropic multilayered system:

$$C\frac{\partial T}{\partial t} = k_x \frac{\partial^2 T}{\partial x^2} + k_y \frac{\partial^2 T}{\partial y^2} + k_z \frac{\partial^2 T}{\partial z^2} + 2k_{xy}\frac{\partial^2 T}{\partial x \partial y} + 2k_{xz}\frac{\partial^2 T}{\partial x \partial z} + 2k_{yz}\frac{\partial^2 T}{\partial y \partial z} \quad \text{(S1-1)}$$

This parabolic partial differential equation can be simplified by doing Fourier transforms to in-plane coordinates and time, $T(x, y, z, t) \leftrightarrow \Theta(u, v, z, \omega)$

$$F(u) = \int_{-\infty}^{\infty} f(x) e^{-i2\pi u x} dx$$

$$\mathbb{F}\left\{\frac{df(x)}{dx}\right\} = i2\pi u F(u)$$

$$\mathbb{F}\left\{\frac{d^2 f(x)}{dx^2}\right\} = -(2\pi u)^2 F(u)$$

$$(iC\omega)\Theta = -4\pi^2 \left(k_x u^2 + 2k_{xy} uv + k_y v^2\right)\Theta + 2i2\pi\left(k_{xz} u + k_{yz} v\right)\frac{\partial \Theta}{\partial z} + k_z \frac{\partial^2 \Theta}{\partial z^2} \quad \text{(S1-2)}$$

Or more compactly,

$$\frac{\partial^2 \Theta}{\partial z^2} + \lambda_2 \frac{\partial \Theta}{\partial z} - \lambda_1 \Theta = 0 \quad \text{(S1-3)}$$

where

$$\lambda_1 \equiv \frac{iC\omega}{k_z} + \frac{4\pi^2 \left(k_x u^2 + 2k_{xy} uv + k_y v^2\right)}{k_z} \quad \text{(S1-4)}$$

$$\lambda_2 \equiv 2i2\pi \frac{\left(k_{xz} u + k_{yz} v\right)}{k_z} \quad \text{(S1-5)}$$

The general solution of Eq. (S1-3) is

$$\Theta = e^{u^+ z} B^+ + e^{u^- z} B^- \quad \text{(S1-6)}$$

where $u^+, u^-$ are the roots of the equation $x^2 + \lambda_2 x - \lambda_1 = 0$:

$$u^{\pm} = \frac{-\lambda_2 \pm \sqrt{(\lambda_2)^2 + 4\lambda_1}}{2} \quad \text{(S1-7)}$$

and $B^+, B^-$ are the complex numbers to be determined.

The heat flux can be obtained from the temperature Eq. (S1-6) and Fourier's law of heat conduction $Q = -k_z (d\Theta/dz)$ as:

$$Q = -k_z u^+ e^{u^+ z} B^+ - k_z u^- e^{u^- z} B^- \tag{S1-8}$$

It would be convenient to write Eqs. (S1-6) and (S1-8) as matrices as

$$\begin{bmatrix} \Theta \\ Q \end{bmatrix}_{n,z} = [N]_n \begin{bmatrix} B^+ \\ B^- \end{bmatrix}_n \tag{S1-9}$$

$$[N]_n = \begin{bmatrix} 1 & 1 \\ -k_z u^+ & -k_z u^- \end{bmatrix} \begin{bmatrix} e^{u^+ z} & 0 \\ 0 & e^{u^- z} \end{bmatrix}_n \tag{S1-10}$$

where $n$ stands for the $n$-th layer of the multilayer system, and $z$ is the distance from the surface of the $n$-th layer.

The constants $B^+, B^-$ for the $n$-th layer can also be obtained from the surface temperature and heat flux of that layer by setting $z=0$ in Eq. (S1-10) and performing its matrix inversion:

$$\begin{bmatrix} B^+ \\ B^- \end{bmatrix}_n = [M]_n \begin{bmatrix} \Theta \\ Q \end{bmatrix}_{n,z=0} \tag{S1-11}$$

$$[M]_n = \frac{1}{k_z (u^+ - u^-)} \begin{bmatrix} -k_z u^- & -1 \\ k_z u^+ & 1 \end{bmatrix} \tag{S1-12}$$

For heat flow across the interface, an interface conductance $G$ is defined. Therefore, the heat flux across an interface can be written as:

$$Q_{n,z=L} = Q_{n+1,z=0} = G \left( \Theta_{n,z=L} - \Theta_{n+1,z=0} \right) \tag{S1-13}$$

From Eq. (S1-13) we also have:

$$\Theta_{n+1,z=0} = \Theta_{n,z=L} - \frac{1}{G} Q_{n,z=L} \tag{S1-14}$$

It is convenient to write Eqs. (S1-13) and (S1-14) as a matrix as

$$\begin{bmatrix} \Theta \\ Q \end{bmatrix}_{n+1,z=0} = [R] \begin{bmatrix} \Theta \\ Q \end{bmatrix}_{n,z=L} \tag{S1-15}$$

$$[R] = \begin{bmatrix} 1 & -1/G \\ 0 & 1 \end{bmatrix} \tag{S1-16}$$

The surface temperature and heat flux can thus be related to those at the bottom of the substrate as

$$\begin{bmatrix} \Theta \\ Q \end{bmatrix}_{n,z=L_n} = [N]_n [M]_n \cdots [R]_1 [N]_1 [M]_1 \begin{bmatrix} \Theta \\ Q \end{bmatrix}_{1,z=0} = \begin{bmatrix} A & B \\ C & D \end{bmatrix} \begin{bmatrix} \Theta \\ Q \end{bmatrix}_{1,z=0} \quad \text{(S1-17)}$$

Applying the boundary condition of zero heat flux at the bottom of the semi-infinite substrate yields $0 = C\Theta_{1,z=0} + DQ_{1,z=0}$. The Green's function $\hat{G}$, which is essentially the surface temperature response due to the applied surface heat flux of unit strength, can thus be solved as

$$\hat{G}(u,v) = \frac{\Theta_{1,z=0}}{Q_{1,z=0}} = -\frac{D}{C} \quad \text{(S1-18)}$$

With the Green's function $\hat{G}$ determined, the detected temperature response is simply the product of $\hat{G}$ and the heat source function in the frequency domain.

## S2: Effect of optical penetration on the surface temperature rise

Throughout this work we have assumed a surface heat source boundary condition, i.e., the laser power was deposited on the very top surface of the first layer. In reality, because of the finite value of the extinction coefficient $\kappa$, the laser beam would penetrate into the material, giving a volumetric heat source. The Beer-Lambert law states that the intensity of the laser beam inside the material would have an exponential decay as $\exp(-z/\delta_p)$, where $\delta_p = \lambda_0/(4\pi\kappa)$ is known as the optical penetration depth or skin depth. Figure S1 (a) shows the cumulative absorption through the depth of Al at a wavelength of $\lambda_0 = 800$ nm, which has a refractive index of $n = 2.67$ and an extinction coefficient of $\kappa = 8.27$. With the slab divided into many sub-layers, the percentage of the amount of heat absorbed by the $i$-th sub-layer with thickness $\Delta z$ at depth $z_i$ could thus be determined as $\beta_i = \frac{1}{\max(\alpha)} \frac{d\alpha(z_i)}{dz} \Delta z$, here $\alpha(z_i)$ is the cumulative absorption at depth $z_i$. When the sub-layers are divided to be sufficiently thin, the volumetric heat source inside a sub-layer can be assumed to be a surface heat source for that sub-layer. A schematic of the sub-divided multilayer model to simulate volumetric heat source is illustrated in Fig. S1 (b). The transfer matrix, instead of being Eq. (S1-17), should thus be replaced as:

$$\begin{aligned}
\begin{bmatrix} \Theta \\ Q \end{bmatrix}_{n,z=L_n} &= [N]_n [M]_n \left\{ \begin{bmatrix} 0 \\ \beta_3 Q_0 \end{bmatrix} + \cdots \left\{ \begin{bmatrix} 0 \\ \beta_3 Q_0 \end{bmatrix} + [R]_2 [N]_2 [M]_2 \left( \begin{bmatrix} 0 \\ \beta_2 Q_0 \end{bmatrix} + [R]_1 [N]_1 [M]_1 \begin{bmatrix} \Theta_1 \\ \beta_1 Q_0 \end{bmatrix} \right) \right\} \right\} \\
&= \{[N]_n [M]_n \cdots [R]_1 [N]_1 [M]_1\} \begin{bmatrix} \Theta_1 \\ \beta_1 Q_0 \end{bmatrix} + \{[N]_n [M]_n \cdots [R]_2 [N]_2 [M]_2\} \begin{bmatrix} 0 \\ \beta_2 Q_0 \end{bmatrix} + \cdots + \{[N]_n [M]_n\} \begin{bmatrix} 0 \\ \beta_n Q_0 \end{bmatrix} \\
&= \begin{bmatrix} A_{n1} & B_{n1} \\ C_{n1} & D_{n1} \end{bmatrix} \begin{bmatrix} \Theta_1 \\ \beta_1 Q_0 \end{bmatrix} + \beta_2 Q_0 \begin{bmatrix} B_{n2} \\ D_{n2} \end{bmatrix} + \cdots + \beta_n Q_0 \begin{bmatrix} B_{nn} \\ D_{nn} \end{bmatrix}
\end{aligned}$$
(S2-1)

Here we have made the following notation:

$$\begin{bmatrix} A_{jk} & B_{jk} \\ C_{jk} & D_{jk} \end{bmatrix} = \{[N]_j [M]_j \cdots [R]_k [N]_k [M]_k\} \qquad \text{with } j \geq k \qquad (S2-2)$$

Applying the same boundary condition $Q_{n,z\to\infty} = 0$ at the bottom of the substrate, we have $0 = C_{n1}\Theta_1 + (D_{n1}\beta_1 + D_{n2}\beta_2 + \cdots + D_{nn}\beta_n)Q_0$. We can thus solve the surface temperature rise of the top layer $\Theta_1$ as

$$\Theta_1 = -\frac{1}{C_{n1}}(D_{n1}\beta_1 + D_{n2}\beta_2 + \cdots + D_{nn}\beta_n)Q_0 \tag{S2-3}$$

An alternative approach to model optical penetration can also be found in Refs. [1, 2], and they yield identical results with the current approach.

Figure S1(c, d) compares the $V_{\text{in}}$ and $V_{\text{out}}$ signals in TDTR experiments simulated by the thermal model assuming surface heat source and volumetric heat source boundary conditions, respectively. Here the simulated data are for 100 nm Al/Si measured by TDTR using a modulation frequency of 5 MHz and a laser spot size of 5 μm. The results suggest that only the $V_{\text{in}}$ signals in the short delay time range 0-10 ps are affected by the heat source boundary condition. The $V_{\text{out}}$ signal, which is due to the heating at the modulation frequency and is analogous to the signals acquired in FDTR experiments, is not affected by the heat source boundary condition. Note that it is the metal transducer layer with a small optical penetration depth that makes it possible to assume a surface heat source boundary condition. For materials with large optical penetration depths, it would be necessary to consider the actual volumetric heat source for a more accurate simulation of the experimental signal, see Ref. 2 for an example.

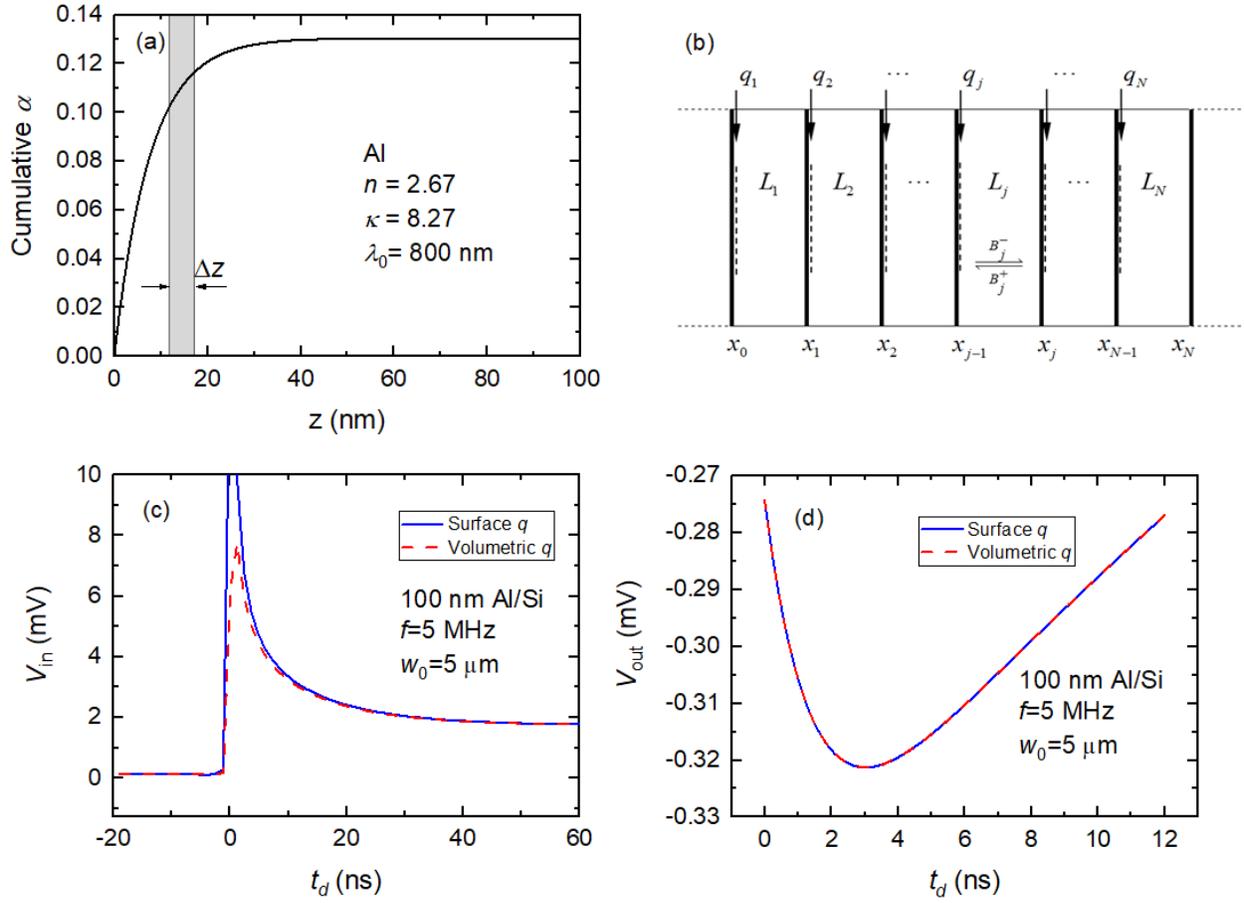

Fig. S1. (a) Cumulative absorption as a function of depth in Al, which has a refractive index of 2.67 and an extinction coefficient of 8.27 at the wavelength 800 nm. (b) General multilayer model for the case of volumetric heat source boundary condition. Each layer has a heat input of $q_j = \beta_j q_0$ on the surface of that layer. (c, d) Simulated in-phase and out-of-phase signals in TDTR experiments for 100 nm Al/Si assuming surface heat source boundary condition (solid curves) and volumetric heat source boundary condition (dashed curves).

# S3: Deriving the correlation (Eq. (18)) for $\theta_{dc}^{pulsed}$

The temperature rise due to the constant offset component of the pulsed laser heating $\theta_{dc}^{pulsed}$ contains two parts: one is a constant component due to the pulse accumulation $\theta_{dc,accum}^{pulsed}$, and the other is a transient component due to the heating by each pulse $\theta_{dc,trans}^{pulsed}$. Obviously, $\theta_{dc,accum}^{pulsed}$ depends on the pulse repetition rate $f_{rep}$, whereas $\theta_{dc,trans}^{pulsed}$ depends on the delay time $t_d$. Here we derive their correlations separately.

We start with the known analytical expression of $\theta_{dc}^{cw} = \xi P_0/\sqrt{2\pi w_0^2 k_z k_r}$. There are many parameters affecting $\theta_{dc,accum}^{pulsed}$. To study the effect of each parameter, we keep all the other factors constant and vary that parameter over a wide range of its possible values, and evaluate the relation between the ratio $\theta_{dc,accum}^{pulsed}/\theta_{dc}^{cw}$ and that parameter. After repeating this process for all the parameters, we find that if we define a factor $x = \left(\dfrac{2w_0}{\sqrt{\dfrac{k_r}{\pi C f_{rep}}}}\right)^{0.36+0.84\xi}$ and plot the ratio $\theta_{dc,accum}^{pulsed}/\theta_{dc}^{cw}$ as a function of $x$, almost all the points collapse into the curve $y = x/(1+x)$, as shown in Fig. S2(a).

Similarly, for $\theta_{dc,trans}^{pulsed}$, we find that if we take the factor $\left(\dfrac{\theta_{dc,trans}^{pulsed}}{\theta_{dc}^{cw}}\right) t_d$ as $y$, and define the $x$ factor as $x = \xi w_0/\sqrt{k_r t_d/\pi C}$, almost all the points collapse into the curve $y = 2 \times 10^{-8} x/(25 + x^2)$, as shown in Fig. S2(b).

Expanding these correlations, we have the expression of $\theta_{dc}^{pulsed}$ as Eq. (18) in the main text.

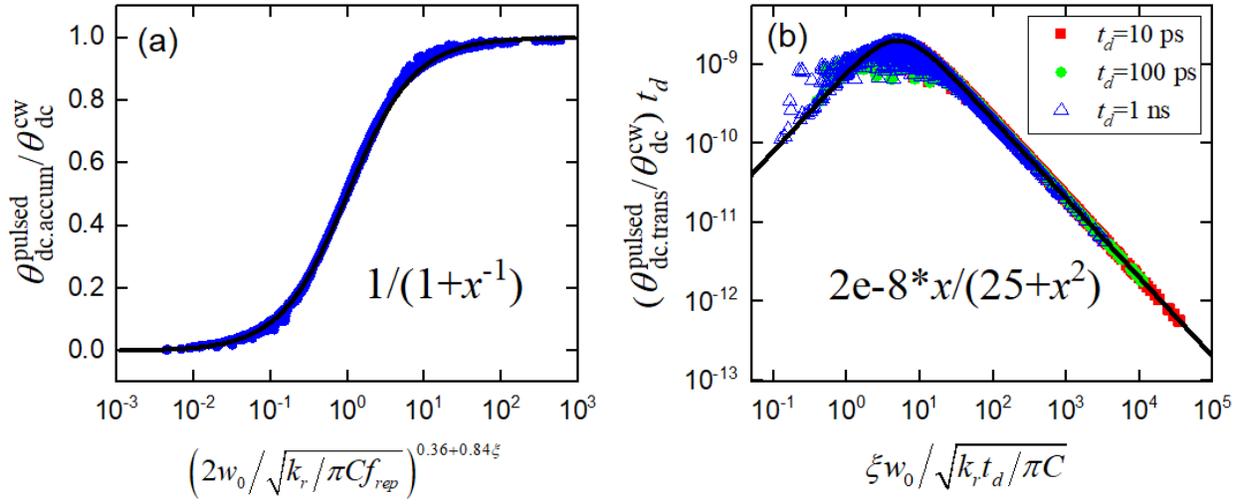

Fig. S2. The key correlated factors determining $\theta_{dc,accum}^{pulsed}$ and $\theta_{dc,trans}^{pulsed}$ in Eq. (18).

# S4: Validation of the correlation (Eq. (18)) for $\theta_{dc}^{pulsed}$

When a semi-infinite solid is heated by a train of unmodulated laser pulses with an averaged heating power of $P_0$, the peak temperature rise $\theta_{dc}^{pulsed}$ is expressed as the first term of Eq. (17) and could be approximated by the analytical correlation Eq. (18) in the main text, i.e.,

$$\theta_{dc}^{pulsed} = P_0 \sum_{n=-\infty}^{\infty} \left[ e^{-\frac{n^2\omega_s^2\tau_p^2}{11.08}} e^{in\omega_s(t-t_0)} \int_{-\infty}^{\infty}\int_{-\infty}^{\infty} \frac{\exp(-\pi^2 u^2 w_x^2/2)\exp(-\pi^2 v^2 w_y^2/2)}{\sqrt{iC(n\omega_s)k_z + 4\pi^2 k_z(k_x u^2 + k_y v^2)}} dudv \right]$$

$$= \theta_{dc,accum}^{pulsed} + \theta_{dc,trans}^{pulsed} \qquad (S4\text{-}1)$$

$$\approx \xi \frac{P_0}{\sqrt{2\pi w_0^2 k_z k_r}} \left[1 + \left(16\frac{k_r}{\pi w_0^2 C}\frac{1}{f_{rep}}\right)^{0.18+0.42\xi}\right]^{-1} + \frac{1.1 P_0}{f_{rep}\pi w_0^2 \sqrt{k_z C t_d}} \left(1 + \frac{25}{\xi^2}\frac{k_r}{\pi w_0^2 C} t_d\right)^{-1}$$

To evaluate the validity of Eq. (18), we generate 1000 random cases, with $k_x, k_y, k_z$ independently and randomly varying in the range of 0.5-5000 W/mK, $C$ varying in the range of 0.001-10 MJ/m³K, and $w_x, w_y$ independently and randomly varying in the range of 1-100 µm, as shown in Fig. S1. The tested anisotropy $k_z/k_r$ and $\alpha/\beta$ have covered a wide range from $10^{-4}$ to $10^4$, as shown in Fig. S3. For each case, we calculate $\theta_{dc}^{pulsed}$ as a function of the time $t$ numerically from the first term of Eq. (17), which is considered as an accurate value. We then calculate $\theta_{dc}^{pulsed}$ for four selected delay time of $t_d = -20, 10, 100,$ and $1000$ ps using the correlation Eq. (18). Note that for $t_d = -20$ ps, $\theta_{dc}^{pulsed}$ was evaluated by the first term of Eq. (18) only. The deviations between the estimated values and the accurate ones, $\eta = \frac{\theta^{estimate} - \theta^{numerical}}{\theta^{numerical}} \times 100\%$, are plotted as a function of $\xi w_0 \sqrt{C f_{rep}/k_r}$ in Fig. S4.

We note that it is not practically useful to estimate the transient temperature rise at a short delay time of 10 ps in TDTR experiments because at such a short delay time the temperature rise is still complicated by factors including the non-equilibrium between hot electrons and phonons in the transducer layer, the deviation between the actual volumetric heat flux due to the optical penetration of the laser beam and the assumed surface heat flux boundary condition, etc. Nevertheless, we can still test our analytical expressions at any delay time within the framework of the simplified thermal model to demonstrate the consistency of the analytical expressions.

Figure S4 shows that the correlation Eq. (18) is accurate with an error of <3% for $t_d = 10$ ps if the dimensionless factor $\xi w_0 \sqrt{Cf_{rep}/k_r} > 1$, which could be met in most cases in TDTR experiments. If $\xi w_0 \sqrt{Cf_{rep}/k_r} < 1$ or if $t_d$ is too large (e.g., $t_d > 1$ ns) so that the three-dimensional heat flow dominates, and the simple expression of Eq. (18) would cause a larger error, but still generally less than 50%.

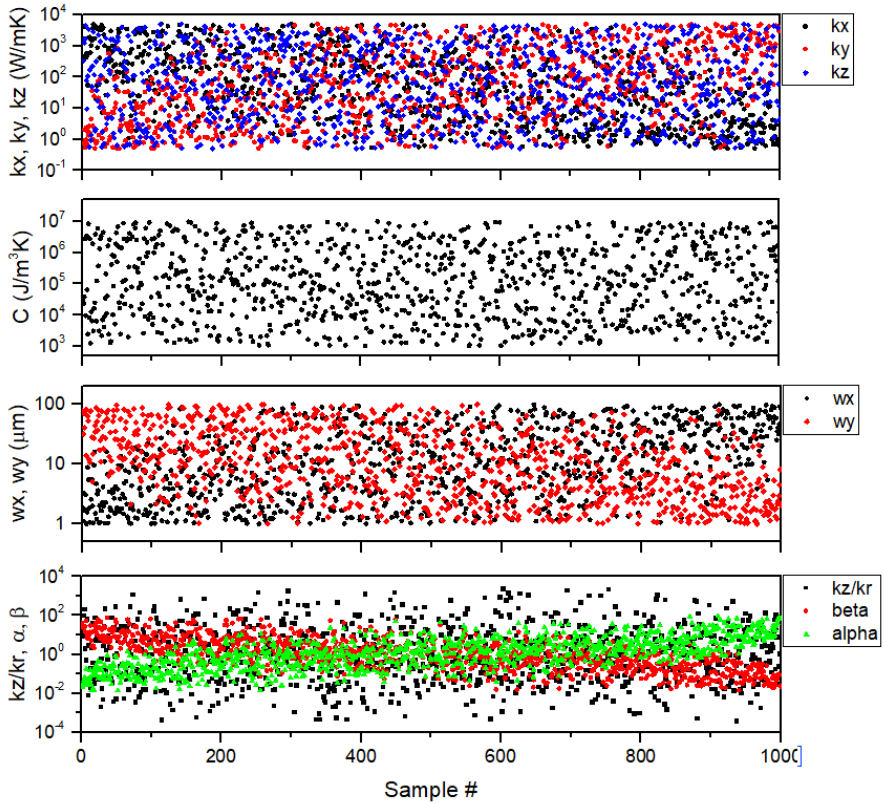

Fig. S3. Parameters of 1000 randomly generated cases for the test of Eq. (18).

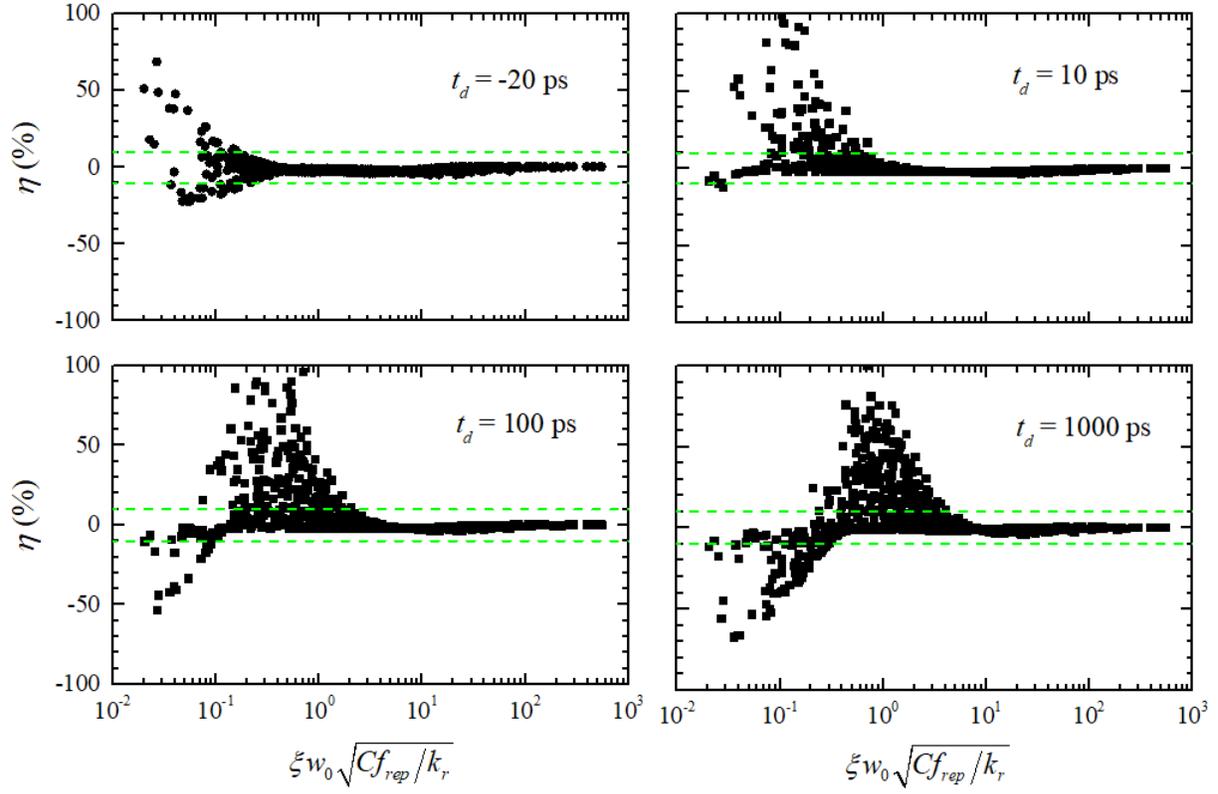

Fig. S4. Deviations of the estimated temperature rise values using Eq. (18) from the accurate numerical values at different delay time of $t_d$ =-20, 10, 100, and 1000 ps.

## S5: The key factor determining the relative amplitudes of $\theta_{dc,accum}^{pulsed}$ and $\theta_{dc,trans}^{pulsed}$

Equation (18) in the main text shows that the temperature rise due to a train of unmodulated laser pulses contains two components, one is a constant value due to the pulse accumulation $\theta_{dc,accum}^{pulsed}$ and the other is a transient value due to the pulse heating $\theta_{dc,trans}^{pulsed}$. The ratio between these two components could be approximated as

$$\theta_{dc,accum}^{pulsed}/\theta_{dc,trans}^{pulsed} \approx \xi \frac{P_0}{\sqrt{2\pi w_0^2 k_z k_r}} \bigg/ \frac{1.1 P_0}{f_{rep} \pi w_0^2 \sqrt{k_z C t_d}} \approx \frac{w_0}{\sqrt{k_r/f_{rep}^2 t_d C}}. \qquad (S5\text{-}1)$$

Therefore, the ratio $w_0/\sqrt{k_r/f_{rep}^2 t_d C}$ is the key factor that determines the relative magnitudes of the pulse-accumulation temperature rise and the transient pulse temperature rise: with $w_0 \gg \sqrt{k_r/f_{rep}^2 t_d C}$ the pulse-accumulation temperature rise would dominate over the transient pulse temperature rise; on the other hand, if $w_0 \ll \sqrt{k_r/f_{rep}^2 t_d C}$, the pulse-accumulation temperature rise would also be much smaller than the transient pulse temperature rise. This conclusion is also verified against 1000 random cases (whose parameters are shown in Fig. S3), with the results shown in Fig. S5.

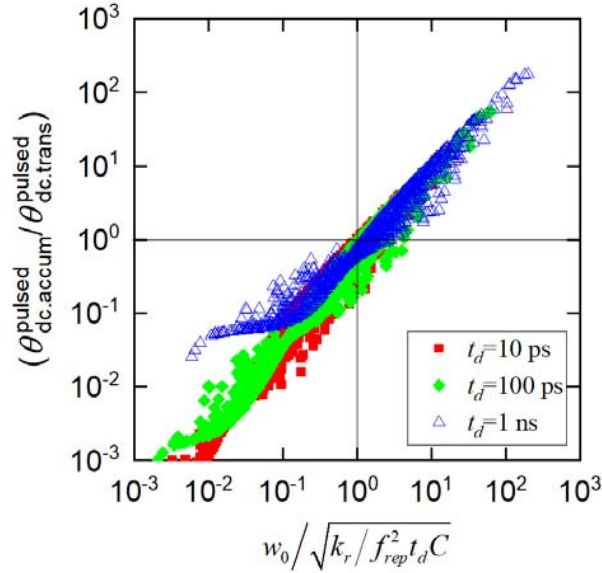

Fig. S5. Test results demonstrate a correlation between the temperature rise rato $\theta_{dc,accum}^{pulsed}/\theta_{dc,trans}^{pulsed}$ and the lateral length scale ratio $w_0/\sqrt{k_r/f_{rep}^2 t_d C}$.

# S6: Validation of the correlations (Eq. (19)) for $\theta_{ac}^{pulsed}$

When a semi-infinite solid is heated by a train of sinusoidally modulated laser pulses with the modulation frequency $\omega_0$ and the amplitude of the modulated heating power as $P_1$, the peak temperature rise $\theta_{ac}^{pulsed}$ is expressed as the second term of Eq. (17) and could be approximated by the analytical correlation Eq. (19) in the main text, i.e.,

$$\theta_{dc}^{pulsed} = P_1 \text{Re}\left\{\sum_{n=-\infty}^{\infty}\left[e^{-\frac{n^2\omega_s^2\tau_p^2}{11.08}} e^{in\omega_s(t-t_0)} e^{i\omega_0 t}\int_{-\infty}^{\infty}\int_{-\infty}^{\infty}\frac{\exp(-\pi^2 u^2 w_x^2/2)\exp(-\pi^2 v^2 w_y^2/2)}{\sqrt{iC(\omega_0+n\omega_s)k_z + 4\pi^2 k_z(k_x u^2 + k_y v^2)}} du dv\right]\right\}$$

$$= \theta_{ac,accum}^{pulsed} + \theta_{ac,trans}^{pulsed}$$

$$\approx \frac{P_1}{P_0}\theta_{dc,accum}^{pulsed}\left[1+4\left(\frac{w_0}{5}\sqrt{\frac{\omega_0 C}{2\pi k_r}}\right)^{0.36+0.84\xi}\right]^{-1}\left[1+\pi\sqrt{\frac{\omega_0}{2\pi f_{rep}}}\right]^{-1} \quad (S6\text{-}1)$$

$$+\frac{P_1}{P_0}\theta_{dc,trans}^{pulsed}\left[1-\left(1+\frac{(\omega_0/2\pi)^{-0.1\log_{10}(4t_d f_{rep})}}{-\log_{10}(4t_d f_{rep})}\right)^{-1}\right]\left(1+150\left(10w_0\sqrt{\frac{\omega_0 C}{2\pi k_r}}\right)^{-0.6-1.4\xi}\right)^{-1}$$

To evaluate the validity of Eq. (19), we select seven modulation frequencies of 10, $10^2$, ..., and $10^7$ Hz. For each frequency, we generate 1000 random cases, with $k_x, k_y, k_z$ independently and randomly varying in the range of 0.5-5000 W/mK, $C$ varying in the range of 0.001-10 MJ/m³K, and $w_x, w_y$ independently and randomly varying in the range of 1-100 μm, similar to the case shown in Fig. S3. For each case, we calculate $\theta_{ac}^{pulsed}$ as a function of the time $t$ numerically from the second term of Eq. (17), which is considered as an accurate value. We then calculate $\theta_{ac,accum}^{pulsed}$ and $\theta_{ac,trans}^{pulsed}$ using the correlation Eq. (19). Three delay time of $t_d = 10, 100,$ and $1000$ ps were chosen to test the correlation of $\theta_{ac,trans}^{pulsed}$.

Figure S6 shows the temperature rises $\theta_{ac,accum}^{pulsed}$ and $\theta_{ac,trans}^{pulsed}$ estimated using correlation Eq. (19), and compared to the accurate numerical values for the 8000 randomly generated cases. Here the heating power was fixed as $P_1 = 1$ mW for all the cases. We can see that the correlation Eq

(19) generally works well. They would have a larger error for large $t_d$, where the heat flow would become highly three-dimensional and would be too complicated to be evaluated by a simple correlation.

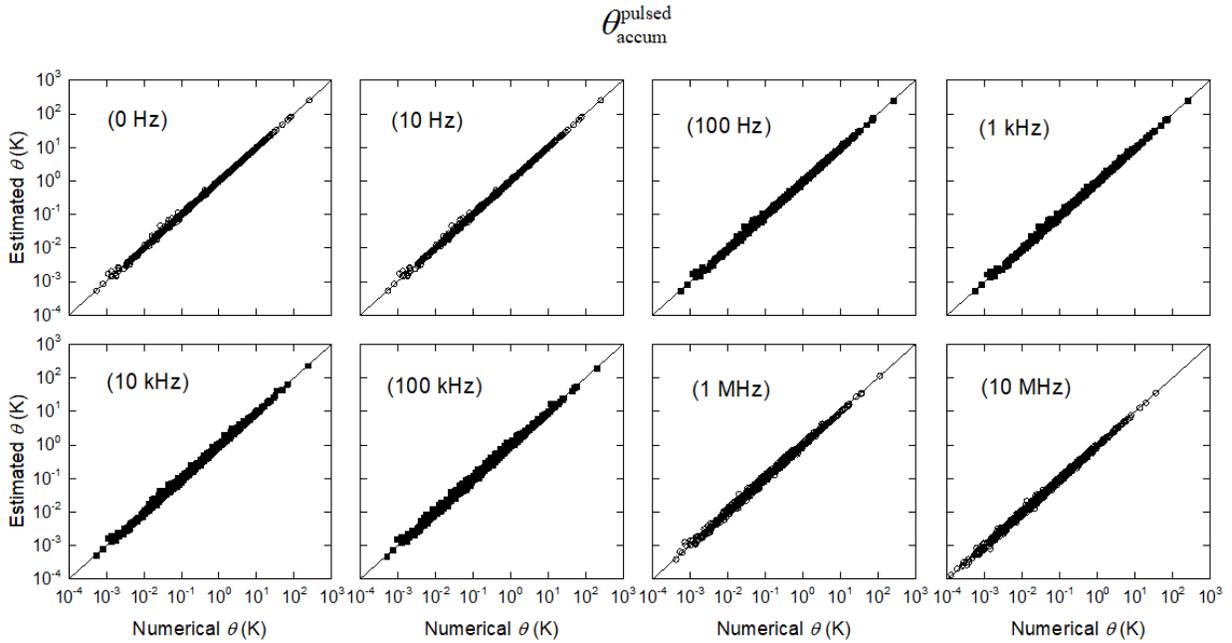

$\theta_{\text{accum}}^{\text{pulsed}}$

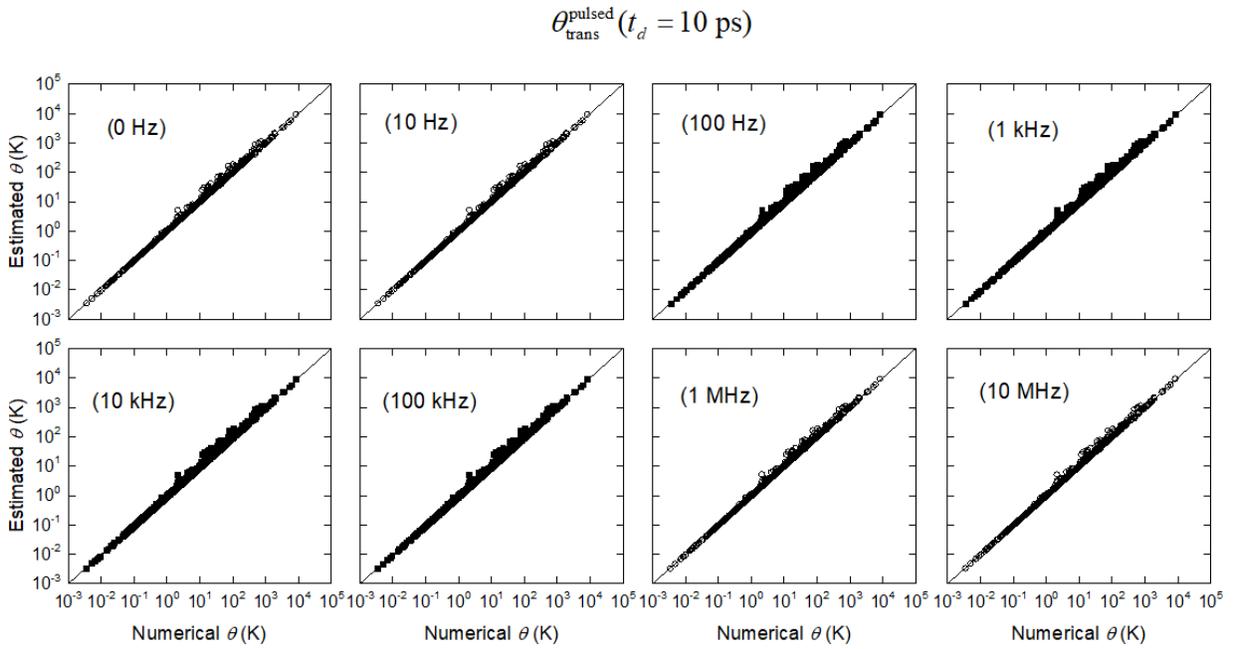

$\theta_{\text{trans}}^{\text{pulsed}}(t_d = 10 \text{ ps})$

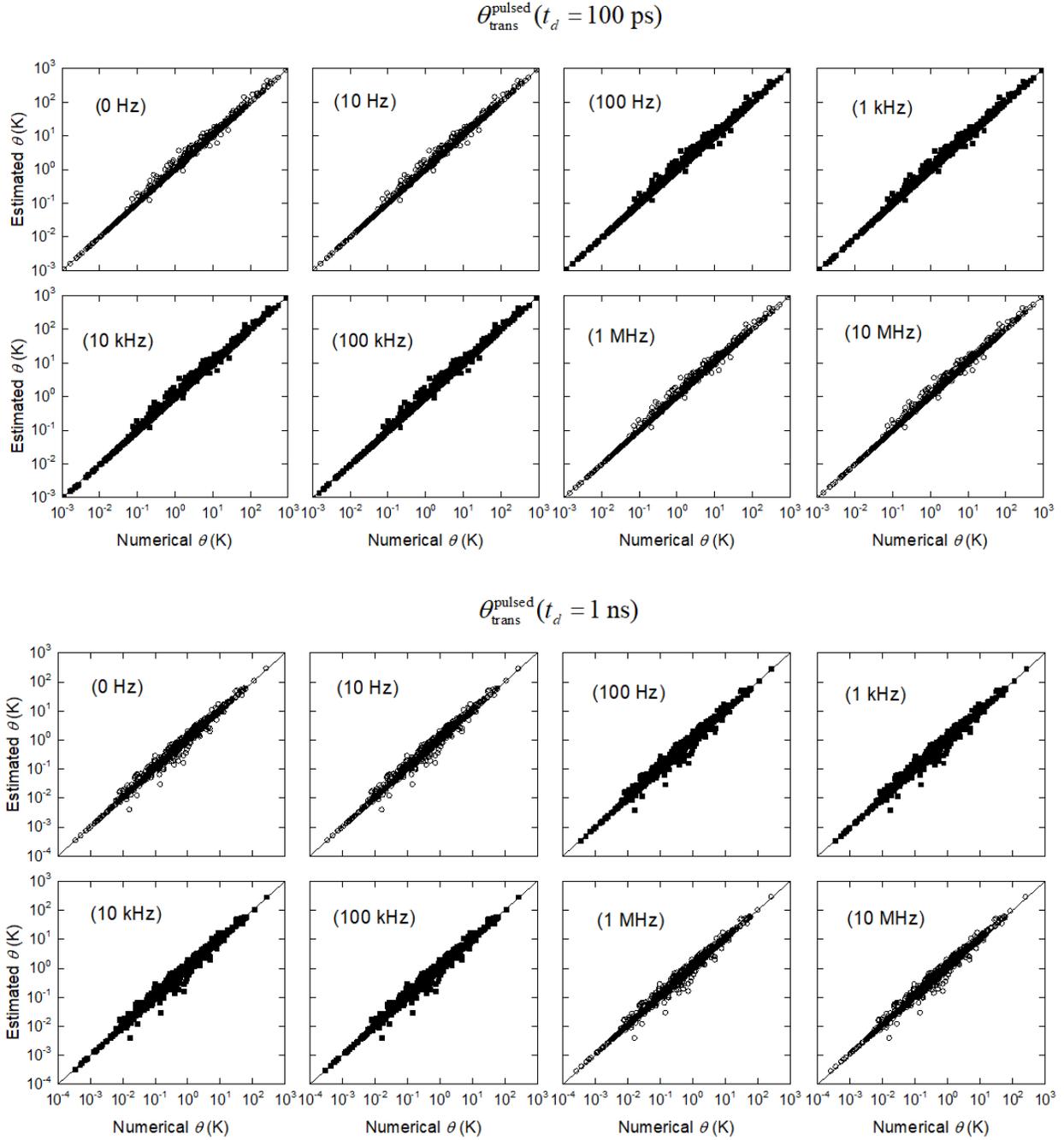

Fig. S6. Estimated temperature rise values using Eq. (19) as compared to the accurate numerical values for modulated pulse heating at different delay time of $t_d$ =-20, 10, 100, and 1000 ps for different modulation frequencies from 10 Hz to 10 MHz.

# S7: A table of parameters for the cases in Fig. 5 in the main text

TABLE S1. Parameters of the cases in Fig. 5 in the main text

| | | Layer 1 | Layer 2 | Layer 3 | |
|---|---|---|---|---|---|
| *Material* | System 1 | Al | SiO$_2$ | Si | |
| | System 2 | Ti | Si | SiO$_2$ | $w_x = 5$ μm |
| $h$ (nm) | System 1 | 100 | 50 | inf | $w_y = 10$ μm |
| | System 2 | 100 | 200 | inf | $P_0 = 10$ mW |
| $k$ (W/mK) | System 1 | 200 | 1.4 | 140 | $P_1 = 10$ mW |
| | System 2 | 10 | 100 | 1.4 | $G = 300$ MW/m$^2$K |
| $C$ (MJ/m$^3$K) | System 1 | 2.4 | 1.67 | 1.6 | |
| | System 2 | 2.35 | 1.6 | 1.67 | |